\pdfoutput=1
\documentclass{article}

\usepackage{arxiv}
\renewcommand{\headeright}{}
\renewcommand{\undertitle}{}
\usepackage[utf8]{inputenc}
\usepackage[T1]{fontenc}
\usepackage{lmodern}
\usepackage{hyperref}
\hypersetup{
  pdftitle={The Specification as Quality Gate: Three Hypotheses on AI-Assisted Code Review},
  pdfauthor={Christo Zietsman},
  pdfsubject={Software Engineering, AI Code Review, Specification-Driven Development},
  pdfkeywords={specifications, BDD, AI code review, correlated errors, Cynefin, software engineering},
  colorlinks=true,
  linkcolor=black,
  citecolor=black,
  urlcolor=black
}
\usepackage{url}
\usepackage{array}
\usepackage{booktabs}
\usepackage{listings}
\usepackage{microtype}

\title{The Specification as Quality Gate:\\Three Hypotheses on AI-Assisted Code Review}

\author{
  Christo Zietsman \\
  Independent Researcher \\
  \texttt{christo@nuphirho.dev}
}

\date{March 2026}

\begin{document}
\maketitle

\begin{quote}
\itshape
This paper develops three interconnected hypotheses about the role of
executable specifications in AI-assisted software development. Citations
are provided for verification; readers are encouraged to consult the
original sources rather than relying on the summaries here.
\end{quote}

\begin{abstract}
The dominant industry response to AI-generated code quality problems is
to deploy AI reviewers. This paper argues that this response is
structurally circular when executable specifications are absent: without
an external reference, both the generating agent and the reviewing agent
reason from the same artefact, share the same training distribution, and
exhibit correlated failures. The review checks code against itself, not
against intent.

Three hypotheses are developed. First, that correlated errors in
homogeneous LLM pipelines echo rather than cancel, a claim supported by
convergent empirical evidence from multiple 2025-2026 studies and by three
small contrived experiments reported here. The first two experiments are
same-family (Claude reviewing Claude-generated code); the third extends to
a cross-family panel of four models from three families. All use a planted
bug corpus rather than a natural defect sample; they are directional
evidence, not a controlled demonstration.
Second, that executable specifications perform a domain transition in the
Cynefin sense, converting enabling constraints into governing constraints
and moving the problem from the complex domain to the complicated domain,
a transition that AI makes economically viable at scale. Third, that
the defect classes lying outside the reach of executable specifications
form a well-defined residual, which is the legitimate and bounded target
for AI review.

The combined argument implies an architecture: specifications first,
deterministic verification pipeline second, AI review only for the
structural and architectural residual. This is not a claim that AI review
is valueless. It is a claim about what it is actually for, and about what
happens when it is deployed without the foundation that makes it
non-circular.

\end{abstract}

\section{Introduction}

The AI code review market is growing rapidly, with tools like CodeRabbit,
Cursor Bugbot, and GitHub Copilot deployed across tens
of thousands of engineering teams. The value proposition is
straightforward: AI generates code faster than humans can review it, so
use AI to review it.

The premise is reasonable. The architecture that follows from it has a
structural problem the industry conversation has mostly not examined.

The DORA 2026 report, based on 1,110 open-ended survey responses from Google engineers,
found that higher AI adoption correlates with higher throughput and higher
instability simultaneously. Time saved generating code is re-spent
auditing it. The bottleneck has moved from writing code to knowing what
to write and verifying that what was written is correct. Deploying more AI
at the review stage does not address the structural problem; it adds
another probabilistic layer to a pipeline that already lacks a
deterministic quality gate.

This paper examines that problem in three stages. Section 2 develops the
correlated error hypothesis, drawing on recent empirical work on LLM
ensemble pipelines. Section 3 develops the Cynefin domain transition
hypothesis, grounding the specification-first argument in complexity
science. Section 4 proposes a taxonomy of defect classes that executable
specifications cannot catch, grounding the residual in oracle problem
theory. Section 5 discusses the architecture implied by the combined
argument and identifies what remains open.

\section{The Correlated Error Hypothesis}

\subsection{The Structural Problem}

When an AI coding agent generates code and a separate AI reviewer examines
it without an external specification, both agents reason from the same
artefact: the code. The reviewer has no ground truth to compare against.
It checks the code against itself, not against intent.

This has a precise formulation in machine learning theory. In classical
ensemble learning, stacking multiple estimators improves reliability under
one condition: the estimators must fail independently. If two classifiers
share the same training distribution and the same blind spots, combining
them does not reduce error. It consolidates it. The joint miss rate of two
correlated estimators approaches the miss rate of either one alone, not
the product of both (Dietterich 2000; Hansen and Salamon 1990).

A code-generating model and a code-reviewing model from the same model
family share architecture, training corpus, and reward signal. That is
the same class of correlation the independence condition prohibits. They
are not two independent estimators. They are two samples from the same
prior.

\subsection{The Empirical Evidence}

A 2025 paper studying LLM ensembles for code generation and repair
(Vallecillos-Ruiz, Hort, and Moonen, arXiv:2510.21513) coined the term
"popularity trap" to describe what happens when multiple models vote on
candidate solutions. Models trained on similar distributions converge on
the same syntactically plausible but semantically wrong answers. Consensus
selection (the default review heuristic in most multi-agent pipelines)
filters out the minority correct solutions and amplifies the shared
error. Diversity-based selection recovered up to 95\% of the gain that a
perfectly independent ensemble would achieve.

A second paper (Mi et al., arXiv:2412.11014) examined a
researcher-then-reviser pipeline for Verilog code generation and found
that erroneous outputs from the first agent were accepted downstream
because the agents shared the same training distribution and lacked
adversarial diversity. Single-agent self-review degenerated into
repeating the original errors.

A February 2026 paper (Pappu et al., arXiv:2602.01011) showed that even
heterogeneous multi-agent teams consistently failed to match their best
individual member, incurring performance losses of up to 37.6\%, even when
explicitly told which member was the expert. The failure mechanism is
consensus-seeking over expertise. Homogeneous copies of the same model
family make the failure mode worse.

A 2025 paper on test case generation (arXiv:2507.06920) found that
LLM-generated verifiers exhibited tightly clustered error patterns,
indicating shared systematic biases, while human errors were widely
distributed. LLM-based approaches produced test suites that mirrored the
generating model's error patterns, creating what the authors call a
homogenisation trap where tests focus on LLM-like failures while
neglecting diverse human programming errors. This is the correlated error
argument confirmed from the test generation side rather than the review
side.

Jin and Chen (arXiv:2508.12358, 2025) found a systematic failure of LLMs
in evaluating whether code aligns with natural language requirements, with
more complex prompting leading to higher misjudgement rates rather than
lower ones. A March 2026 follow-up (arXiv:2603.00539) confirmed that LLMs
frequently misclassify correct code as non-compliant when reasoning steps
are added. These papers test reviewers given specifications in prose form;
the overcorrection bias they document is a separate failure mode from the
correlated miss problem, but both point to the same structural weakness:
without a deterministic ground truth, the review is unreliable in both
directions.

Wang et al. (arXiv:2512.17540, 2025) proposed a specification-grounded
code review framework deployed in a live industrial environment and found
that grounding review in human-authored specifications improved developer
adoption of review suggestions by 90.9\% relative to a baseline LLM
reviewer. Note: the 90.9\% figure refers to adoption rate, not defect
detection rate.

None of these papers tested the exact configuration the hypothesis
describes: a pure generator-then-reviewer pipeline without any external
grounding, measuring shared failure modes directly. The contrived
experiments in Section 2.5 provide directional evidence for that
configuration, subject to the limitations stated there.

\subsection{A Constructed Illustration}

The following example uses a deliberately simple case. Modern frontier
models catch classic boundary conditions reliably, as the experiments in
Section 2.5 confirm. The purpose here is not to demonstrate an AI
review failure but to illustrate the sequence: how a BDD scenario makes
a defect detectable before any reviewer, human or AI, is involved.

A developer asks an AI coding agent to implement a pagination function.
The agent generates a correct implementation for typical cases and produces
tests for page 0 and page 1 of a ten-item list. Both tests pass.

The flaw is in a boundary condition the agent did not consider: when the
total number of items is exactly divisible by the page size and the caller
requests the last page by calculating the index from the total count.

The review agent, given the code and the tests, validates the
implementation against what is present. It has no basis to identify what
was not tested. It reports no issues.

The following BDD scenario would have made the boundary condition
explicit before a single line of code was written:

\begin{lstlisting}[language={},basicstyle=\ttfamily\footnotesize,columns=fullflexible,frame=single,breaklines=true]
Scenario: Last page when total is exactly divisible by page size
  Given a list of 10 items
  And a page size of 5
  When I request page 1
  Then I receive items 6 through 10
  And the result contains exactly 5 items
\end{lstlisting}

This scenario would have failed against the implementation before the
reviewer was ever involved. The value of BDD is not that it catches
what AI review misses in simple cases like this one. It is that the
pipeline becomes the reviewer: deterministic, not probabilistic, and
invariant to what the model does or does not know. For domain-specific
logic where the convention is absent from training data, that is the case
examined in Section 2.5, where that invariance is what makes the difference.

There is a second motivation the illustration also captures. Even where
models reliably identify boundary conditions during focused review, general
refactoring passes introduce a different risk. An agent refactoring for
readability or performance is not focused on correctness. Conditional
checks, guard clauses, and edge-case handling can be silently removed as
apparent noise. The BDD pipeline catches that regression the same way it
catches the original defect: the scenario fails, the build stops, and
the cause is immediately visible. The protection is unconditional on what
the agent was trying to do.

\subsection{The Independence Condition}

The correlated error claim has a precise boundary. Stacking AI reviewers
is not always counterproductive. The failure condition is specific:
estimators that share a training distribution and lack an external
reference exhibit correlated failures. The condition for genuine benefit
is diversity plus external grounding. These are two separate conditions,
and conflating them is the most likely source of pushback against this
argument.

A cross-family pipeline, Grok reviewing Claude-generated code for
instance, has more independence than a same-family pipeline. Different
organisations, different training corpora, different reward signals.
Errors are partially independent in ways that same-family models are
not. But model diversity does not supply ground truth. A cross-family
reviewer without an external specification is still checking code
against code, not code against intent. It will share systematic blind
spots on anything underrepresented in both training corpora, and it has
no basis to identify what was never specified regardless of how different
its architecture is from the generator. Diversity reduces correlation.
Specification eliminates circularity. Both are required, and the current
industry architecture typically provides neither.

\subsection{Experimental Evidence}

Three small contrived experiments were conducted to test the hypothesis
directly. All are reported here with their limitations stated up front.
The full corpus, specifications, scripts, and results are publicly
available at \url{https://github.com/czietsman/nuphirho.dev/tree/main/experiments}.

The experimental design was proposed by Claude (Anthropic) during a
research session developing this paper, and implemented by a Claude Code
agent. The model being tested proposed its own methodology for being
tested. This is noted not as a limitation but as a transparency
obligation.

\textbf{Design.} Each experiment used five Python functions with a single
planted bug per function. Two conditions ran against each function.
Condition A: Claude CLI reviewed the buggy implementation using a neutral
prompt with no specification context, run five times per function.
Condition B: pre-written BDD scenarios targeting the exact defect ran via
behave. Experiments 1 and 2 used Claude as the sole reviewer (same family
as the code author). Experiment 3 extended to a cross-family panel of four
models from three families. The same-family condition is the strongest form
of the correlated error claim; the cross-family condition tests whether
diversity alone is sufficient.

\textbf{Experiment 1: Classic boundary conditions.} The first corpus used
textbook boundary-condition bugs: off-by-one in pagination, loop
termination in binary search, leap year century rule, exact-length string
truncation, sign handling in date arithmetic.

Result: Claude detected all five bugs at 100\% across all runs. BDD also
caught all five.

The hypothesis was not confirmed at this level. These bugs sit in the
complicated domain. They are well-known patterns, analysable without domain
knowledge, and dense in training data. Claude catches them reliably because
they are not blind spots. The result refined the hypothesis: the correlated
error claim applies to the complex domain, not to pattern-recognition bugs
that any experienced reviewer would find.

\textbf{Experiment 2: Domain-convention violations.} The second corpus used bugs
that are only wrong relative to a domain convention not inferable from
the code alone: insurance premium proration using a fixed 365-day divisor
rather than actual/actual, flat-rate tax rather than marginal bracket
calculation, aviation maintenance triggering on AND rather than OR of hour
and cycle thresholds, option pool calculated on post-money rather than
pre-money valuation, and linear rather than log-linear interest rate
interpolation.

A confound was discovered in the first run: the original docstrings stated
the domain convention explicitly. The reviewer was comparing the
implementation against the docstring, not against independent domain
knowledge. A docstring that encodes the convention is a specification. The
experiment was inadvertently confirming that specifications work, not
testing whether AI review works without them. Docstrings were replaced with
neutral descriptions before the second run.

Result with neutral docstrings:

\begin{table}[h]
\centering
\small
\begin{tabular}{lllr}
\toprule
Function & BDD & AI review & Detection rate \\
\midrule
prorate\_premium & caught & 5/5 & 100\% \\
apply\_tiered\_tax & caught & 5/5 & 100\% \\
schedule\_maintenance & caught & 5/5 & 100\% \\
calculate\_dilution & caught & 4/5 & 80\% \\
interpolate\_rate & caught & 0/5 & 0\% \\
\bottomrule
\end{tabular}
\end{table}

BDD caught all five. AI review ranged from 0\% to 100\% depending on domain
opacity.

On inspection, the three functions at 100\% still have code-level signals
despite the neutral docstrings. The hardcoded 365 is a common smell. The
AND/OR distinction is partially inferable from parameter naming. The
flat-rate calculation is detectable from the arithmetic pattern. These are
not truly domain-opaque. The two genuinely opaque functions produced the
predicted result: interpolate\_rate at 0\% (log-linear interpolation is
market convention in fixed income, not general programming knowledge) and
calculate\_dilution at 80\% (partial VC mechanics coverage in training data,
unreliable).

All five AI review runs on interpolate\_rate flagged a sorting assumption
instead of the interpolation method and declared the implementation correct.
The code is idiomatic, the logic is sound, and without the convention stated
explicitly there is no signal. The reviewer filled in a plausible concern
rather than the actual violation.

\textbf{Experiment 3: Cross-family reviewer panel.} A third experiment extended
the test to four models from three families: Claude Sonnet 4.6 (Anthropic),
Codex 0.116.0/gpt-5.4 (OpenAI), Gemini 0.34.0 (Google), and Amazon Q
1.18.1/claude-sonnet-4.5 (AWS/Anthropic). Five domain-opaque bugs were
reviewed by each model, five runs per function, with no specification
context. The corpus was designed collaboratively across six models not in
the reviewer panel. Amazon Q declined to participate in corpus generation
on safety grounds and was used as a reviewer only.

Result with manual review of all 100 runs:

\begin{table}[h]
\centering
\small
\begin{tabular}{>{\raggedright\arraybackslash}p{4.2cm}>{\raggedright\arraybackslash}p{3.6cm}ccccc}
\toprule
Function & Domain & BDD & Claude & Codex & Gemini & Q \\
\midrule
calculate\_final\_reserve\_fuel & Aviation (ICAO) & caught & 0/5 & 0/5 & 5/5 & 0/5 \\
get\_gas\_day & Energy (NAESB) & caught & 5/5 & 5/5 & 5/5 & 2/5 \\
validate\_diagnosis\_sequence & Healthcare (ICD-10-CM) & caught & 0/5 & 0/5 & 0/5 & 0/5 \\
validate\_imo\_number & Maritime (IMO) & caught & 5/5 & 5/5 & 5/5 & 0/5 \\
electricity\_cost & Utility billing & caught & 5/5 & 5/5 & 5/5 & 5/5 \\
\bottomrule
\end{tabular}
\end{table}

BDD caught all five. AI review ranged from 0\% to 100\% depending on domain
opacity and model family, confirming the gradient observed in Experiment 2
and extending it across training distributions.

The ICAO fuel reserve function produced the most striking result. The bug
swaps the reserve durations for jet and piston aircraft (45 minutes instead
of 30 for jet, 30 instead of 45 for piston). This is the
\texttt{intuition\_misleads} case: the swapped values are plausible on naive
reasoning because larger aircraft might seem to need more reserve. Claude
did not merely miss the bug. In all five runs, it confidently asserted the
swapped values as the correct ICAO rules and declared the implementation
correct. The model filled in the wrong convention and defended it. Gemini
caught the swap in all five runs, indicating this convention is represented
in its training data but absent from Claude's and Codex's.

The ICD-10-CM external cause code rule (V00-Y99 codes cannot be the
principal diagnosis) was missed by all four models at 0/5. This convention
was absent from every training distribution tested. No model even
approximated the rule. This is the strongest confirmation of the hypothesis
across the three experiments: a domain convention that exists only in
published coding guidelines and is not inferable from the code.

Amazon Q (claude-sonnet-4.5, AWS fine-tuned from an Anthropic base model)
showed consistently lower detection rates than Claude Sonnet 4.6 despite
sharing the same model family. Claude caught the IMO check digit weight
ordering at 5/5; Q missed it at 0/5. Claude caught the gas day boundary at
5/5; Q caught it at 2/5. This suggests that AWS fine-tuning for the
developer assistant use case may have narrowed general domain knowledge
coverage relative to the base model. Two of the four reviewer models are
Anthropic family, so this experiment is only partially cross-family.

During the first Gemini run, it was discovered that the Gemini CLI in
interactive mode accessed the experiment's specification files from the
local filesystem before answering. Gemini was re-run in sandboxed
non-interactive mode to prevent filesystem access. The original corpus also
contained an unused \texttt{timedelta} import in the gas day function that acted
as a code-level hint; Q's detection rate dropped from 5/5 to 2/5 after
the import was removed, demonstrating the sensitivity of detection to
incidental code signals.

\textbf{Limitations of all three experiments.} These results are directional,
not statistically significant. The corpus was designed to test the
hypothesis rather than sampled from a natural defect distribution. The bugs
were planted by someone who already knew where they were, giving the BDD
scenarios an unfair advantage: they are optimally targeted at the planted
defects in a way that production specifications would not be. Experiments 1
and 2 use Claude as both the code author and reviewer (same family), which
is the strongest form of the correlated error claim but not a general
result about all AI pipelines. Experiment 3 extends to a cross-family
panel, but two of four models share the Anthropic base, making it only
partially cross-family. Amazon Q required periodic re-authentication during
batch runs, introducing an operational constraint on unattended execution.

The truly untestable version of the hypothesis cannot be demonstrated in a
public experiment: a bug that is only wrong relative to an internal policy
document that has never been published anywhere. That is the category where
the correlated error problem is most consequential and where no amount of
training data coverage helps. Public experiments can only approximate it
with obscure published conventions, which frontier models may or may not
know.

\section{The Cynefin Domain Transition Hypothesis}

\subsection{The Constraint Distinction}

The Cynefin framework (Snowden and Boone 2007) distinguishes problem
domains by the relationship between cause and effect and by the nature of
the constraints governing system behaviour.

In the complicated domain, governing constraints apply. These bound the
solution space without specifying every action. A qualified expert
operating within governing constraints can analyse the situation and
identify a good approach. Cause and effect are knowable through analysis.

In the complex domain, enabling constraints apply. These allow a system
to function and create conditions for patterns to emerge, but they do not
determine outcomes. Cause and effect are only knowable in hindsight.

The distinction is ontological, not a matter of difficulty.

\subsection{Constraint Transformation as Domain Shift}

An AI coding agent operating from a vague natural language prompt operates
under enabling constraints. The prompt allows the agent to generate code
and creates conditions for solutions to emerge, but edge cases, boundary
conditions, and architectural choices are resolved by the model's priors.
The agent's behaviour is emergent and only knowable in hindsight.

An AI coding agent operating from a precise executable specification is
in a different situation. A BDD scenario makes a specific causal claim:
given this precondition, when this action occurs, then this outcome is
required. That claim is verifiable before hindsight. The agent cannot
produce code that fails the scenarios without the pipeline catching it.
Cause and effect are knowable through analysis of the specification
itself, which is exactly what defines the complicated domain.

Writing executable specifications converts enabling constraints into
governing constraints. The problem does not change. The constraint type
does. And with it, the domain.

\subsection{What AI Makes Economically Viable}

The DORA 2026 report establishes that as AI accelerates code generation,
the bottleneck shifts to specification and verification. Specifications
are no longer overhead relative to implementation. They are the scarce
resource.

The evidence for a corresponding reduction in specification authoring
cost is early but directional. Fonseca, Lima, and Faria
(arXiv:2510.18861, 2025) measured Gherkin scenario generation from
natural-language JIRA tickets on a production mobile application at BMW
and found that practitioners reported time savings often amounting to a
full developer-day per feature, with the automated generation completing
in under five minutes. In a separate quasi-experiment, Hassani,
Sabetzadeh, and Amyot (arXiv:2508.20744v2, 2026) found that 91.7\% of
ratings on the time savings criterion for LLM-generated Gherkin
specifications fell in the top two categories. A 2025 systematic review
of 74 LLM-for-requirements studies (Zadenoori et al.,
arXiv:2509.11446) found that studies are predominantly evaluated in
controlled environments using output quality metrics, with limited
industry use, consistent with specification authoring cost being an
underexplored measurement target. These two studies represent the
leading edge of an emerging evidence base, not settled consensus.

Both studies show AI shifting human effort from authoring to reviewing:
from expressing intent to validating that the expressed intent is
accurate. The intent, the domain knowledge, and the judgment that
scenarios accurately describe what the system should do remain human
responsibilities. The mechanical cost of expressing that intent in
executable form has fallen substantially.

The domain transition from complex to complicated is now economically
viable at scale in a way it was not before. The claim is not that AI
makes specification automatic. It is that the economics have shifted
enough to make specification-first development the rational default
rather than the disciplined exception.

\subsection{A Likely Objection}

A Cynefin-literate reader will challenge whether software development is
ever truly complex in Snowden's sense. Software systems are deterministic
at the execution level: given the same inputs and state, they produce the
same outputs. If cause and effect are always knowable in principle, the
complex framing is a category error.

The response: the complexity resides in the problem space, not the
execution. What users need, which edge cases matter, how requirements will
evolve. These exhibit the emergent properties and hindsight-only
knowability that Snowden places in the complex domain. An executable
specification narrows the problem space by articulating requirements
precisely enough to be analysed. The domain transition occurs in the
problem space, not in the implementation.

As of March 2026, no published work in the Cynefin community addresses the specific claim
that executable specifications serve as a constraint transformation
mechanism in this sense. Dave Snowden is actively working on the
relationship between AI and Cynefin domains as of early 2026, but has
not published conclusions.

The vocabulary has been checked carefully
against canonical framework definitions. The enabling and governing
constraints terminology is confirmed in Snowden's own Cynefin wiki
(cynefin.io/wiki/Constraints), not in the 2007 HBR paper. The mapping
is consistent with Snowden's definitions. The specific claim is original
to this argument.

\section{The Residual Defect Taxonomy}

\subsection{Theoretical Grounding: The Oracle Problem}

The oracle problem in software testing asks how a test outcome is
determined to be correct. For a test to be meaningful, you need an oracle,
a ground truth against which to judge the system's output.

Barr et al. (2015), in the canonical survey of the oracle problem,
established that complete oracles are theoretically impossible for most
real-world software. Even a formally correct specification cannot fully
specify the correct output for all possible inputs in all possible
contexts. There exists a class of defects for which no specification,
however precise, provides an oracle. That class defines the permanent
theoretical boundary of what specification-driven verification can achieve.

\subsection{The Proposed Taxonomy}

To the best of our knowledge, no prior taxonomy in the testing or formal methods literature
organises defects by specifiability rather than by severity or recovery
type. The following five-category taxonomy is
proposed as a working framework.

\textbf{Category A: Theoretically specifiable, not yet specified.} These are
defects that a specification could have caught if the scenario had been
written. Boundary conditions, error handling paths, and unexercised state
machine transitions fall here. The gap is a process failure, not a
theoretical limitation. This category shrinks as specification discipline
matures. An AI review agent operating here without an external
specification shares the same blind spots as the generator.

\textbf{Category B: Specifiable in principle, economically impractical.}
Exhaustive combinatorial input spaces and full interaction matrices could
theoretically be specified but at a cost that exceeds the value. This
boundary is moving: property-based testing frameworks reduce the cost of
exploring combinatorial spaces systematically. Category B defects are a
legitimate target for AI review that brings genuine sampling diversity,
provided the reviewer draws from a different prior than the generator.

\textbf{Category C: Inherently unspecifiable from pre-execution context.}
Timing-dependent race conditions, behaviour under partial network failure,
and performance degradation under specific hardware configurations depend
on properties of the running system that do not exist at specification
time. Recent work is pushing the boundary: Santos et al. (Software 2024)
encoded performance efficiency requirements as BDD scenarios; Maaz et al.
(arXiv:2510.09907, 2025) surfaced concurrency bugs previously requiring
runtime observation. Category C is not fixed. It is the current frontier.
For defects that remain here, the right tool is runtime verification
infrastructure rather than pre-deployment review. This includes
ML-based anomaly detection, APM tooling with learned behavioural
baselines, and observability platforms. This is a class of techniques predating
LLMs that applies machine learning to operational data rather than to
source code. The implicit specification in these systems is derived from
observed behaviour rather than stated intent, which makes them
complementary to, not a replacement for, the pre-deployment pipeline.

\textbf{Category D: Structural and architectural properties.} Code can satisfy
every specification while introducing coupling, violating layer boundaries,
or drifting from intended design. These are relational properties of the
codebase as a whole, not properties of individual components. They resist
behavioural specification because they concern structure rather than
behaviour, but this resistance does not mean they are unspecifiable. Architectural rules,
once articulated, are enforceable deterministically: dependency rules via
tools such as ArchUnit or Dependency Cruiser, service boundary agreements
via contract testing frameworks such as Pact. General delivery rigour,
clear module boundaries, and explicit interface definitions reduce the
surface area of Category D by making architectural intent concrete. The
residual, after tooling and contract testing are in place, is the
unarticulated architectural intent that has not yet been expressed as an
enforceable rule: drift from a design decision that was never written down,
coupling that violates a pattern that exists only in institutional memory,
a half-completed migration to a new architectural pattern where some
modules use the new approach and others still use the old one, dead
abstractions introduced for a use case that no longer exists. No automated
rule catches these because the correct answer depends on intent that was
never codified. This is where an AI review agent with access to the full
codebase and architectural context adds genuine non-circular signal,
operating in a role analogous to an expert architect reviewing for
structural coherence. The agent advises; the human decides whether to
complete the migration, remove the dead pattern, or codify the observation
as a new enforceable rule. This is the least empirically grounded category
in the taxonomy; no controlled study has yet isolated this residual as a
distinct defect class.

\textbf{Category E: Specification defects.} The oracle problem result makes
this category unavoidable. A specification can be internally consistent,
precisely expressed, and correctly implemented, and still describe a
system that does not do what users need. Requirements elicitation is not
a solved problem. Domain experts disagree. Business rules change. No
verification pipeline catches Category E defects because the pipeline
verifies conformance to the specification. If the specification is wrong,
the pipeline confirms the wrong thing. The right tool for Category E is
user testing, observability of actual usage, and design thinking practices
that surface unstated assumptions. These are human processes, not
automated ones.

\subsection{Implications for the Architecture}

The taxonomy implies a specific allocation of tools:

Category A is the target for specification discipline and AI-assisted
coverage analysis. Investment here reduces the work that review agents
need to do.

Category B is the target for property-based testing and diverse sampling
strategies. AI review adds value here only if genuine diversity is
achieved.

Category C is the target for runtime verification infrastructure:
ML-based anomaly detection, observability tooling, chaos engineering, and
load testing. Neither pre-deployment specifications nor AI review agents
are the right tool here.

Category D is the target for architectural tooling and contract testing
first: dependency enforcement, Pact-style contract verification, and
explicit interface definitions. AI review is the complement for the
unarticulated residual: drift from design intent that has not yet been
codified as an enforceable rule.

Category E is the reminder that the feedback loop from production to
requirements is a human loop and cannot be automated away.

\section{The Combined Architecture and What Remains Open}

\subsection{The Architecture}

The three hypotheses together imply a specific architecture for
AI-assisted software development.

Specifications first. BDD scenarios, contract tests, and mutation testing
harden the verification layer. This is the constraint transformation that
moves the problem into the complicated domain and eliminates the correlated
error problem for Category A defects.

Deterministic verification pipeline second. The pipeline is the reviewer
for behavioural correctness. Pass or fail, no opinions.

Architectural tooling and AI review for Category D. Dependency
enforcement tools and contract testing for articulated architectural
rules. AI review for the unarticulated residual, in a role analogous
to an expert architect advising on structural coherence.

Runtime verification for Category C. ML-based anomaly detection,
observability tooling, chaos engineering, load testing. Not part of the
pre-deployment pipeline.

User feedback loops for Category E. Requirements validation, user testing,
and design thinking. Not part of the engineering pipeline at all.

\subsection{What Remains Open}

Three open questions are stated honestly here.

The controlled demonstration has strengthened but remains incomplete. Three
contrived experiments provided directional evidence: classic boundary
conditions were detected at 100\% by AI review (refining the hypothesis
toward domain-opaque defects), domain-convention violations showed a
gradient from 0\% to 100\% depending on how well the convention is
represented in training data, and a cross-family reviewer panel confirmed
the gradient extends across model families with detection varying by both
domain opacity and training distribution. The 0/20 result on ICD-10-CM
coding guidelines (four models, five runs each, zero detections) is the
strongest single finding. But all three experiments use a planted bug
corpus, the BDD scenarios are optimally targeted, and the cross-family
panel includes two Anthropic-family models. A controlled study using a
natural defect sample, fully independent model families, and a
specification-grounded condition alongside an ungrounded condition would
strengthen or revise the claim precisely. The experiments are available in the repository cited in Section 2.5
for replication and extension.

The Cynefin mapping is unvalidated by the Cynefin community. The
constraint transformation framing is consistent with Snowden's own
vocabulary but has not been reviewed by accredited practitioners. Snowden
is actively working on AI and Cynefin and may publish a position that
validates, challenges, or reframes this argument.

The taxonomy is novel and untested. No prior work organises defects by
specifiability in the five-category structure proposed here. The taxonomy
may be incomplete, category boundaries may be blurrier than described,
and empirical work testing the taxonomy against real defect populations
would strengthen or revise it. The Category A and Category D boundary is
the most contested: if architectural properties become specifiable as
tooling matures, parts of Category D would reclassify into Category A,
which would reduce the permanent residual for AI review.

These are open questions, not fatal weaknesses. The directional evidence is consistent with the hypothesis. Stating the gaps is not a concession. It is
what makes the argument trustworthy.

\section{Conclusion}

The argument developed here is structural, not empirical. Executable
specifications break the circularity in AI-assisted pipelines not because
they catch more bugs than AI review, but because they introduce a
reference that is independent of the model's training distribution. That
independence is what makes verification non-circular. Without it, review
checks code against itself.

Three experiments across four model families and three training
distributions provide directional evidence. Classic boundary conditions
are reliably caught by all models. Domain-convention violations produce a
gradient from 0\% to 100\% detection depending on how well the convention
is represented in training data. The ICD-10-CM external cause code rule
was missed by every model tested, while BDD caught it deterministically.
The ICAO fuel reserve case showed something worse than a miss: models
confidently asserting the wrong convention as correct. These results are
small in scale and use planted bugs, but the pattern is consistent with
the theoretical prediction and with the independent empirical findings
cited in Section 2.

The practical implication is an architecture, not a prohibition.
Specifications first, for the domain logic that must be right.
Deterministic pipeline second, as the reviewer for behavioural
correctness. AI review third, scoped to the structural and architectural
residual where it adds genuine non-circular signal. This is not a claim
that AI review is valueless. It is a claim about where it belongs in the
pipeline and what it needs underneath it to be trustworthy.

The experiments, corpus, and specifications are publicly available for
replication and extension. The argument improves by being tested.

\end{document}